\documentclass{article}

\usepackage[preprint, nonatbib]{neurips_2023}

\usepackage[utf8]{inputenc} 
\usepackage[T1]{fontenc}    
\usepackage{hyperref}       
\usepackage{url}            
\usepackage{booktabs}       
\usepackage{amsfonts}       
\usepackage{nicefrac}       
\usepackage{microtype}      
\usepackage{xcolor}         

\title{Artificial Intelligence:\\ Arguments for Catastrophic Risk}

\author{%
  Adam Bales\thanks{Equal contribution. Paper forthcoming at \emph{Philosophy Compass}.}\\
  University of Oxford\\
  \And
  William D'Alessandro$^{*}$\\
  University of Oxford\\
  \And
  Cameron Domenico Kirk-Giannini$^{*}$ \\
  Rutgers University--Newark\\
}

\begin{document}

\maketitle

\begin{abstract}
Recent progress in artificial intelligence (AI) has drawn attention to
the technology's transformative potential, including what some see as
its prospects for causing large-scale harm. We review two influential
arguments purporting to show how AI could pose catastrophic risks. The
first argument --- the \emph{Problem of Power-Seeking} --- claims that,
under certain assumptions, advanced AI systems are likely to engage in
dangerous power-seeking behavior in pursuit of their goals. We review
reasons for thinking that AI systems might seek power, that they might
obtain it, that this could lead to catastrophe, and that we might build
and deploy such systems anyway. The second argument claims that the
development of human-level AI will unlock rapid further progress,
culminating in AI systems far more capable than any human --- this is
the \emph{Singularity Hypothesis}. Power-seeking behavior on the part of
such systems might be particularly dangerous. We discuss a variety of
objections to both arguments and conclude by assessing the state of the
debate.
\end{abstract}

\section{Introduction}

Artificial intelligence (AI) has arrived. Today's AI systems display
various impressive capabilities: they handily defeat the best humans at
games of strategy (Schrittwieser et al. 2019), score highly on
standardized tests (OpenAI 2023a), help design novel drugs (Jumper et
al. 2021, Bran et al. 2023), create dazzling images via models like
Dall-E 3, write and debug code via models like Github Codex, and
converse compellingly with humans via models like GPT-4.

These capabilities are likely to progress rapidly in coming years as
competition intensifies and hardware and algorithms improve. Some
leading AI companies explicitly aim at creating general-purpose
human-level AI.\footnote{E.g., Google's DeepMind: ``Our long term aim is
  to solve intelligence, developing more general and capable
  problem-solving systems, known as artificial general intelligence''
  (\href{https://www.deepmind.com/about}{{https://www.deepmind.com/about}}).
  And OpenAI: ``We believe our research will eventually lead to
  artificial general intelligence, a system that can solve human-level
  problems''
  (\href{https://openai.com/research/overview}{{https://openai.com/research/overview}}).
  If this optimism among technical experts surprises some readers
  familiar with the skeptical arguments of Hubert Dreyfus (1965, 1972),
  it is worth pointing out that contemporary machine learning has moved
  beyond many of the paradigms against which Dreyfus's criticisms were
  leveled.} Assuming this goal is achieved, developers will face
strategic and economic incentives not to stop at human-level
capabilities. We may soon\footnote{How soon? In a large 2023 survey of
  published AI researchers, superhuman general-purpose AI was expected
  by 2047 with 50\% likelihood (Grace et al. 2024). Meanwhile, as of
  January 2024, the forecasting site Metaculus gives 2032 as the
  aggregated expected date for human-level AI, based on over 2500
  predictions
  (\href{https://www.metaculus.com/questions/5121/date-of-artificial-general-intelligence/}{{https://www.metaculus.com/questions/5121/date-of-artificial-general-intelligence/}}).
  For a comparison and synthesis of a number of estimated timelines, see
  Wynroe et al. (2023). Of course, predicting long-term technological
  developments is no trivial task even for experts, and the historical
  record of AI timeline estimates suggests that developments in the
  field may be especially difficult to foresee.} find ourselves sharing
the world with a new kind of highly capable thinking thing, trained but
not necessarily understood or fully controlled by us.

A growing number of philosophers and others fear that future AI
technology might pose a catastrophic risk to humanity (Bostrom 2014, Ord
2020, Vold \& Harris 2021). In this article, we explore some of the
arguments behind such worries.\footnote{A companion article examines
  proposals for developing AI safely.} In contrast with discussion of
pandemics, nuclear weapons, and other global threats, the AI risk debate
turns on distinctive issues in decision theory, ethics, and philosophy
of mind. For this reason, the debate has been shaped by philosophers
from its beginnings.

Our coverage here is limited in several ways. First, we primarily
discuss arguments for \emph{catastrophic} risk, involving globally
devastating harm or threats to human existence, autonomy or potential.
(This isn\textquotesingle t to deny that major non-catastrophic harms
from AI are also worth taking seriously.) Second, we focus on risk
claimed to be \emph{intrinsic} to certain kinds of AI systems, rather
than risk associated with human error or malicious misuse.\footnote{For
  discussion of biochemical misuse risks, see D'Alessandro et al. 2023.
  For general taxonomies of AI risk, see Yampolskiy 2015, Zwetsloot \&
  Dafoe 2019 and Hendrycks et al. 2023.} (This isn\textquotesingle t to
say that the latter kinds of risks are smaller, but they
won\textquotesingle t be our focus here.) Finally, lacking space to
cover every form of risk in detail, we concentrate on one widely
discussed idea: the notion that some advanced AI systems are likely to
function as agents pursuing goals, and as a result, are likely to engage
in dangerous resource-acquiring, shutdown-avoiding, and
correction-resisting behavior.

Section 2 outlines this \emph{Problem of Power-Seeking.} Section 3 then
discusses two claims about AI goals (the \emph{Instrumental Convergence
Thesis} and the \emph{Orthogonality Thesis}) that are relied upon in
making the case for the Problem of Power-Seeking. Assessing these claims
is crucial for assessing the broader argument. In section 4, we turn to
the \emph{Singularity Hypothesis}, according to which rapid
improvements, culminating in capabilities far beyond our
species\textquotesingle{} level, will follow on the heels of human-level
AI. If this hypothesis is true, it provides one reason to expect the
sort of supercapable systems that can lead to the Problem of
Power-Seeking. So assessing this hypothesis helps us to determine how
concerned we should be by power-seeking AI systems.

It's worth noting that there's little consensus about the likelihood of
catastrophe from the sources we discuss. A small but vocal contingent of
pessimists takes human extinction to be the default outcome of advanced
AI (cf. Yudkowsky 2023). Other theorists take catastrophe to be unlikely
but worth taking seriously, on the grounds that even small chances of
disaster should be understood in detail and reduced if possible. (For
instance, Carlsmith (2022) estimates that an AI catastrophe is
\textgreater10\% likely by 2070.) Still others question whether the
risks rise even to this level of significance (Thorstad 2022, §6.2,
Thorstad 2023). In general, the reader would do well to view the
arguments below as aiming at plausibility rather than very high
probability or proof.

\section{The Problem of Power-Seeking}

Existing AI systems have already caused significant harms. Pedestrians
have been killed by self-driving cars, for instance, and AI-generated
images have aided incendiary misinformation campaigns (Klee \& Ramirez
2023, Sanger \& Myers 2023). However, while such outcomes are tragic,
they don\textquotesingle t constitute catastrophes on the scale
discussed above. To pose such a risk, AI systems would likely need much
greater power to affect the world. One prominent worry suggests that
this is just what might happen: under plausible assumptions, some AI
systems will seek power, successfully obtain it and go awry, potentially
bringing about catastrophe (Yudkowsky 2008, Omohundro 2008, Bostrom
2012, 2014, Carlsmith 2022, forthcoming). This is the \emph{Problem of
Power-Seeking.}

We can spell out the case for this concern by considering four
questions.

First, \textbf{why might AI systems seek power?} The standard argument
appeals to the \emph{Instrumental Convergence Thesis}, which in rough
terms holds that certain subgoals are useful for achieving a wide
variety of final goals (Omohundro 2008, Bostrom 2012). Such subgoals
might include self-preservation, self-improvement, and resource
acquisition, among others. The Instrumental Convergence Thesis seems to
suggest that many AI systems will develop these subgoals, which we can
describe as power-seeking subgoals, and so suggests that AI systems will
seek power. We discuss this thesis further in §3.

Second, \textbf{why might power-seeking AI successfully acquire power?}
One reason is that some expect us to develop \emph{superintelligent AIs}
with radically greater cognitive powers than the smartest humans. Such
systems might be able to engage in sophisticated misinformation
campaigns, develop new military technologies, profit from the stock
market, and so on. Given such advantages, if superintelligent AI systems
seek power, some will plausibly acquire it (Bostrom 2014).

Despite the terminology, this argument doesn\textquotesingle t rely on a
unified notion of general intelligence. Nor does it assume that AIs will
become conscious, acquire personhood, or possess humanlike beliefs and
desires. Instead, what matters is that the systems are highly capable at
various tasks that are useful for gaining power (Carlsmith forthcoming).
A more apt label for this notion might be \emph{supercapability}.

Supercapability provides one reason AI systems might acquire power;
there are at least two others. One is \emph{supernumerosity}: since
copying software is cheap and easy, the population of capable AIs could
quickly grow large. If these systems cooperated, sheer numbers might
allow them to seize substantial collective power (Karnofsky 2022).
Further, humans might voluntarily relinquish significant power to
increasingly capable systems, because these systems can carry out useful
tasks. For example, militaries might hand control of drones to AI.

Third, \textbf{why might AI systems acquiring power lead to
catastrophe?} If AI systems seek power on large scales, this might lead
to conflict with humans as competition arises for resources and
influence. For example, AIs might seek to disempower humans to limit our
interference with their goals and might seize control of resources we
rely on in order to promote their own ends (Bostrom 2014, p. 141). If
these systems have acquired sufficient power, the outcomes of such
behaviors could be catastrophic for humans.

Fourth, \textbf{why would we develop and deploy AI systems that posed
catastrophic risks?} For a start, even if these risks were real and
there was mounting evidence for this, some people might not take the
danger seriously, just as some are unconcerned by climate change.
Further, if AI proves useful and economically valuable, competition
between AI companies and between governments will likely ensue
(Carlsmith 2022, §5). Competitive environments encourage rapid progress
that might not involve adequate attention to safety. Finally,
there\textquotesingle s the possibility of \emph{deceptive alignment},
which involves systems appearing safe during development but becoming
dangerous when they\textquotesingle re deployed in the world (Ngo et al.
2023, §4.2; Bostrom 2014, pp. 142--145). In such cases, we might not
realize the risk until it\textquotesingle s too late.

This concludes our overview of the Problem of Power-Seeking. We now
consider two objections.

First, the above arguments envisage AI systems with at least human-level
cognitive capabilities that behave in ways that strike us as capricious
and morally alien, pursuing unlimited power in the service of arbitrary
goals. This is surprising. Perhaps it's also implausible. As AI systems
develop cognitive capabilities at a similar level to our own, it's
possible their motivations will also converge on something like ours
(see Müller and Cannon 2021). Catastrophic power-seeking might then
be rare. Against this line of thought, \emph{the Orthogonality Thesis}
holds that arbitrarily high levels of intelligence can be combined with
more or less any final goals (Bostrom 2012). If this were true, a system
could be superintelligent without being driven by concerns that are
conducive to human flourishing. We discuss these issues further in
§3.

According to a second objection, power-seeking worries ignore the fact
that humans create AI systems and so control their form. We can design
systems which won't cause large-scale harm, and we can implement
regulations to ensure dangerous systems aren't deployed (cf. Pinker
2015). Indeed, safety concerns and economic incentives point in the same
direction here: unleashing catastrophe on the world would be bad for
business, so even profit-driven companies will be motivated to invest in
safety.

These suggestions are reasonable in principle, but it's unclear that
they'll be easy to implement in practice. Ensuring that AI systems do
what we want requires ruling out misaligned behavior contrary to
designers' intentions.\footnote{For discussions of the concept of
  alignment, see Gabriel 2020, Carlsmith 2022 §4.1.} If misaligned
behavior proves difficult to foresee prior to deployment, as some
suspect it will, then well-meaning developers and regulators may lack
the tools to avert catastrophe.

This \emph{alignment problem} is at the core of many worries about
catastrophic risk from AI. To get a sense of why it might prove
difficult for designers to foresee and rule out misaligned behavior,
consider the story of King Midas---or stories of golems and
genies---where a simple wish is fulfilled in unexpected and undesirable
ways. Part of the point of these stories is that wishes, when
interpreted literally and without reference to our broader assumptions
and desires, can easily go awry in ways we didn\textquotesingle t
anticipate. A supercapable AI system whose motives or understanding of
the world differ even slightly from ours might resemble such dangerous
wish-fulfillers (Muehlhauser and Helm 2012).

More concretely, we can see how misalignment might arise by considering
AI systems developed via \emph{reinforcement learning}, a central
technique underpinning contemporary AI. In reinforcement learning, a
designer doesn\textquotesingle t specify how a system is to carry out a
given task; instead, the system learns how to do so via a training
process. This involves the system being run on a task and then receiving
feedback on how it performed, in the form of a reward signal. The system
is then modified, based on this reward, so that it will better carry out
the task. With enough iterations, this process can lead to highly
competent systems.

Given this approach, there are at least two ways that misaligned
behavior can arise: reward misspecification and goal misgeneralization
(Ngo et al. 2023). \emph{Reward misspecification} involves a system
accidentally being rewarded for undesirable behavior (Krakovna et al.
2020).\footnote{Some terminology: \emph{reward hacking} and
  \emph{specification gaming} refer to behaviors that take advantage of
  reward misspecification to achieve high reward in an undesirable way.}
This happens because it can be hard to specify a reward that captures
exactly the desired behavior. For example, one system designed to win a
boat racing video game was rewarded for achieving in-game points, a
seemingly reasonable choice (Clark and Amodei 2016). However, rather
than winning, the resulting system steered the boat in circles in a way
that allowed it to continually acquire points. Here, the reward was
misspecified.

Even in the absence of misspecification, \emph{goal misgeneralization}
can arise (Shah et al. 2022, Langosco et al. 2022). This phenomenon
arises because, in complex environments, reinforcement learning systems
cannot be trained on every situation they might face. This raises the
possibility that any goals developed might lead to desirable behavior in
the situations encountered in training but lead to undesirable behavior
in novel situations. This is goal misgeneralization. For example, Shah
et al. (2022) trained an AI system to collect virtual apples and avoid
being attacked by monsters. The system learned to collect shields, which
helped with the latter task. Yet it continued to collect shields even
when placed in environments without monsters, where acquiring shields
was a waste of effort. This system\textquotesingle s behavior
generalized poorly to novel situations.

Reward misspecification and goal misgeneralization represent two ways
that AI systems could develop unwanted and potentially dangerous
behavioral tendencies. These might be difficult to anticipate prior to
deployment, so safety-minded intentions might not suffice to prevent
harm.

To summarize: AI risk theorists maintain that we have grounds to think
AI systems might seek and acquire power in a way that leads to
catastrophe, and grounds to think we might deploy such systems anyway.
This is the Problem of Power-Seeking.

\section{AI Goals}

The above argument appealed to two claims relevant to the nature of AI
goals: the Instrumental Convergence Thesis and the Orthogonality Thesis.
We\textquotesingle ll now examine these.

We start with the Instrumental Convergence Thesis, the claim that
certain resource-acquiring, self-improving and shutdown-resisting
subgoals are useful for achieving a wide variety of final goals. Given
this thesis, we might expect sufficiently advanced goal-directed AI
systems to pursue such subgoals.

Instrumental Convergence is motivated by the observation that agents
possessing more material resources, foresight, factual knowledge, and
persistent influence on the world are typically better positioned to get
what they want in many domains (Omohundro 2008, Bostrom
2012).\footnote{These assets are less helpful for some goals, such as
  the goal of doing nothing. This is why Instrumental Convergence is
  framed in terms of a ``wide variety'' of final goals.} For example,
acquisition of resources like money is useful for achieving many goals.
Consider an AI system with the goal of proving mathematical theorems.
This system could use money to hire research assistants or to pay for
computing infrastructure with which to search the space of possible
proofs. Similar motivations can be given for the remaining
subgoals.\footnote{For attempts to formally prove that various behaviors
  are instrumentally convergent, see Turner et al. 2021 and Turner and
  Tadepalli 2022.}

The subgoals comprising the Instrumental Convergence Thesis form a
heterogeneous group; they need not stand or fall as genuinely universal
together. Accordingly, some skeptics have focused their criticisms on
particular members of the set. Salib (forthcoming) argues against
self-improvement as a convergent subgoal, on the grounds that unimproved
AIs may be unable to control the goals of their improved successors, and
may therefore view self-improvement as a risk to be avoided. Meanwhile
Gallow (forthcoming) evaluates the case for various potential subgoals,
reaching the mixed conclusion that some are plausibly instrumentally
convergent while others are not (see also Goertzel 2015).

Alternatively, one might raise a wholesale objection to the Instrumental
Convergence Thesis, as Goertzel does when he suggests that the
thesis\textquotesingle s intuitive plausibility relies on a tendency to
anthropomorphize (Goertzel 2015).\footnote{Another possibility: humans
  might deliberately give AI systems power-seeking goals. If so then the
  problem of power-seeking doesn\textquotesingle t rely on Instrumental
  Convergence.}

We turn now to the Orthogonality Thesis: the claim that arbitrarily high
levels of intelligence can be combined with almost any final goal.
Bostrom's case for this thesis rests on a conception of intelligence as
skill at instrumental reasoning, or ``search{[}ing{]} for instrumentally
optimal plans and policies'' (2012, 73).

Müller and Cannon (2021) accept that Orthogonality might hold for
intelligence in this instrumental sense. They argue, however, that AI
systems would pose a threat to humanity only if they possessed a richer
sort of general intelligence, consisting partly in the ability to
reflect on and modify one's goals. According to Müller and Cannon, the
Orthogonality Thesis is false if interpreted in terms of this latter
notion of intelligence, because a highly capable generally intelligent
system would seek rational and ethical coherence (and so might not
retain any arbitrary initial goal). Of course, both of these claims rest
on controversial assumptions. As an empirical matter, history seems to
offer vivid illustrations of actors threatening large subsets of
humanity without undergoing the kind of transformative self-examination
Müller and Cannon describe; if this combination of general intelligence
and dangerous goals is common for us, why think it's unlikely for AI
systems?

In addition to arguments targeting each thesis in isolation, there are
also holistic considerations relevant to both. We\textquotesingle ll
discuss two of these.

First, Goertzel (2015) notes that questions of safety will be decided by
the specific goals that future AI systems actually develop. These will
likely be constrained---by human intentions, technological limitations,
training materials and procedures, and so on---rather than entirely
arbitrary.

This observation might seem to challenge the relevance of both theses.
Ultimately, what matters is not whether subgoals are useful for
achieving \emph{a wide variety of} final goals, but whether
they\textquotesingle re useful for achieving future systems' actual
goals, which may or may not be typical of goals in general.\footnote{For
  work on whether present-day systems display the behaviors predicted by
  Instrumental Convergence, see Baker et al. 2020, Perez 2022, OpenAI
  2023b, §2.9.} So Instrumental Convergence might be true but irrelevant
for the real cases of interest in our world. Likewise, what matters is
not whether high levels of intelligence can be combined with \emph{most}
final goals, but whether intelligence is compatible with future systems'
actual goals. If not, Orthogonality will be similarly irrelevant.

These observations suggest that power-seeking arguments would be
improved by paying greater attention to the particular types of goals
that future AI systems are most likely to pursue. (The same advice also
applies to some of these arguments' critics: for instance, Gallow's
negative results on Instrumental Convergence assume that an AI
system\textquotesingle s desires are sampled randomly from the space of
all possible desires.) On the other hand, if there are strong reasons to
believe that the appearance of constraint is misleading and future AI
goals might well be essentially arbitrary, these reasons deserve to be
stated more clearly than they have been.

A second holistic point is that Instrumental Convergence and
Orthogonality both rely on the assumption that AI systems will have
goals (and goal-pursuing abilities) of some kind. This assumption may
deserve further scrutiny.\footnote{The question isn\textquotesingle t
  whether human developers of AI will have goals or whether a reward
  signal will be used in training but whether systems will themselves
  have goals (Ngo et al. 2023, §3).}

In its favor, Carlsmith (2022, §3) argues that goal-based architectures
are very useful: systems that have goals and competently aim to achieve
them will be well placed to carry out complex tasks in the real
world.\footnote{Carlsmith\textquotesingle s discussion of whether AI
  systems will be agentic planners is roughly equivalent to our question
  of whether these systems will have goals.} If this is right, then
human designers might deliberately give AI systems the structure needed
for goals (as was tried, for instance, with Auto-GPT, an AI assistant
that augments ChatGPT with internet access and various interactive
capabilities).\footnote{There\textquotesingle s currently a great deal
  of interest in agentized language models, which might suggest that
  systems with goals are likely to become the norm in the future. See,
  for example, Wang et al. 2023, Goldstein and Kirk-Giannini 2023.}
Alternatively, the usefulness of goals might mean that, if we train a
machine learning system to competently carry out a complex task, goal
pursuit might emerge during training, even without us intentionally
creating this structure (Carlsmith 2022, §3.3, Ngo et al. 2023, §3.2).
However, Drexler (2019) pushes back against these views, arguing that it
is both possible and desirable to create advanced AI systems that
don\textquotesingle t have goals in the relevant sense.

A second class of arguments for expecting goal-pursuing AI systems are
so-called coherence arguments (Omohundro 2007, §3, Omohundro 2008,
Yudkowsky 2019). In simple terms, these arguments hold that AI systems
will be open to undesirable forms of manipulation unless they behave as
\emph{expected utility maximizers}, which can be viewed as certain sorts
of goal-driven agents. Given the further claim that sufficiently
advanced AI systems won\textquotesingle t be open to undesirable
manipulation, we get the conclusion that these systems will be expected
utility maximizers. Hence, they\textquotesingle ll have goals in a
relevant sense. However, Thornley (2023) and Bales (forthcoming) have
pushed back against these arguments, in part by noting that the
connection between avoiding manipulation and maximizing expected utility
is less straightforward than simple versions of the coherence argument
assume.\footnote{See also Drexler 2019, §6.}

All of these debates---about the Instrumental Convergence Thesis, the
Orthogonality Thesis, and whether AI will be goal-driven---strike us as
live and worth further philosophical attention.

\section{The Singularity Hypothesis}

The Singularity Hypothesis is the hypothesis that there will be a period
of rapid recursive improvement in the capabilities of AI systems
following the point at which AI systems become able to contribute to
research on AI. It bears on questions of catastrophic risk in so far as
the result of such an ``intelligence explosion'' could be supercapable
artificial systems which, if not aligned with human interests, might
pose a threat to humanity.

Dialectically, the Singularity Hypothesis makes contact with the Problem
of Power-Seeking at two points: first, misaligned agential AI systems
with instrumentally convergent reasons to engage in self-improvement
might trigger a singularity-type recursive process; second, human actors
seeking to improve the capabilities of AI systems could initiate a
singularity-type recursive process which gives rise to misaligned
supercapable AI systems. Thus it is conceivable that an ``intelligence
explosion'' could be either the \emph{result} or the \emph{cause} of
power-seeking behavior in AI systems.

The two most extended treatments of the Singularity Hypothesis from a
philosophical perspective are due to Chalmers (2010) and Bostrom (2014),
and these will be our focus in what follows.\footnote{Other sources
  include Good (1951, 1959, 1962, 1965), Moravec (1988), Vinge (1993),
  Bostrom (1998, 2003), Kurzweil (2005), Hanson (2008), Muehlhauser and
  Salamon (2012), and Loosemore and Goertzel (2012).}

Chalmers (2010) summarizes the case for the Singularity Hypothesis as
follows:

\begin{quote}
1. There will be AI (before long, absent defeaters).

2. If there is AI, there will be AI+ (soon after, absent defeaters).

3. If there is AI+, there will be AI++ (soon after, absent defeaters).

4. So: There will be AI++ (before too long, absent defeaters).
\end{quote}

Here, AI is understood as human-level artificial intelligence, AI+ is
understood as artificial intelligence that surpasses most humans, and
AI++ is understood as intelligence that is ``far greater than human
level'' (2010, 11).

Chalmers motivates premise 1 by suggesting that if evolution could
produce human-level intelligence, there can be no in-principle barrier
to humans themselves producing the same level of intelligence
artificially. The rapid advancement of the capabilities of AI systems
over the past decade's ``deep learning revolution'' further bolsters the
plausibility of this first premise.

When it comes to premise 2, Chalmers suggests that if we create AI using
an ``extendible method'' --- a method that could, if further developed,
yield more intelligent systems --- we should expect to be able to
produce AI+ soon after. Not all methods of creating AI are plausibly
extendible. As Chalmers points out, however, even if AI is not initially
achieved using an extendible method, it could subsequently be used to
discover an extendible method. This would suffice for the truth of
premise 2.

The idea behind premise 3 is that the recursive improvement that will
follow from the rise of AI+ will soon lead to AI++. Chalmers motivates
the idea of recursive improvement by appealing to what he calls the
\emph{proportionality thesis}: the thesis that ``increases in
intelligence (or increases of a certain sort) always lead to
proportionate increases in the capacity to design intelligent systems''
(2010, 21). Given the proportionality thesis, once we reach an AI+ with
the ability to design a system more intelligent than itself, we know
that the system it designs will be able to design a system still
\emph{more} intelligent, and so on until AI++ is achieved.

Bostrom's (2014) version of the argument appeals to interplay between
two factors: \emph{optimization power} --- the amount of effort applied
to improving AI systems, weighted by its quality --- and the
\emph{recalcitrance} of the problem of improving them.\footnote{Bostrom
  writes about optimization power and recalcitrance as though they are
  continuous quantities representable by the real numbers but is
  explicit that this way of framing things is only a heuristic (2014,
  65). We believe the charitable way to read Bostrom is as committed to
  the weakest structural assumptions about intelligence, optimization
  power, and recalcitrance that make modeling future changes in the
  intelligence of AI systems using his equations predictively useful.
  Identifying what these structural assumptions are is an interesting
  problem for future work.} When optimization power is great and
recalcitrance is low, Bostrom suggests, we should expect AI systems to
improve rapidly. On this way of framing the issue, the key question in
considering the Singularity Hypothesis is how we should expect
optimization power and recalcitrance to change as we move from existing
AI systems toward supercapable ones.

Bostrom offers a number of considerations supporting the idea that
recalcitrance is likely to remain constant or decrease in the future,
while optimization power is likely to increase. Even if it proves
difficult to improve the algorithms implemented by artificial systems,
for example, Bostrom argues that this recalcitrance will likely be
offset by the comparative ease of improving such systems by investing in
more hardware and giving them access to more content. At the same time,
we should expect a rapid increase in optimization power due to growing
investment of resources like capital and human researchers and the fact
that, as AI systems become increasingly capable, they themselves will be
able to contribute increasing amounts of optimization power. Indeed,
Bostrom suggests that after the event he calls ``crossover'' --- the
moment when most of the optimization power applied to the problem of
improving AI systems comes from AI systems themselves --- we should
expect the capabilities of AI systems to undergo rapid improvement
because ``any increase in the system's capability {[}will translate{]}
into a proportional increase in the amount of optimization power being
applied to its further improvement'' (2014, 74). Thus Bostrom, like
Chalmers, thinks about the singularity through the lens of a
proportionality thesis.\footnote{While the arguments of Bostrom and
  Chalmers remain the most influential among academic philosophers,
  readers may also be interested in two recent works which adopt a
  ``compute-centric framework'' for estimating how long it will take AI
  systems to reach human-level capabilities (Cotra 2020, Davidson 2023).
  This approach begins by estimating how much computing power would be
  required to train a system with a given level of capability, and then
  constructs a model of when we\textquotesingle re likely to use this
  much computing power to train a system. One way of thinking about
  compute-centric models which we find helpful is as spelling out in
  more detail how the relationship between optimization power and
  recalcitrance is likely to evolve in the future.}

Both Chalmers and Bostrom are aware that the fact that a process of
recursive improvement is possible does not entail that it will actually
occur. Chalmers considers two broad classes of circumstances which might
prevent a singularity-type recursive process from resulting in
superintelligent AI systems. First, \emph{situational defeaters} include
natural or manmade disasters, resource limitations, and other conditions
which might interrupt the process of recursive improvement. Second,
\emph{motivational defeaters} include disinclination on the part of
either humans or AI systems to initiate or maintain the process of
recursive improvement, as well as active attempts to prevent or
interrupt that process. Bostrom discusses similar possibilities in the
context of exploring how recalcitrance might evolve in the future.

From the perspective of concern about catastrophic risks to humanity,
many of the situational defeaters Chalmers considers may strike us as
cold comfort. For example, a natural or manmade disaster severe enough
to permanently arrest technological progress toward superintelligent AI
would itself likely be a global catastrophe.

At the same time, there are reasons to think that the kinds of
motivational defeaters Chalmers considers are unlikely permanently to
prevent a singularity-type outcome. There are strong military and
business incentives for state and corporate actors to develop
increasingly sophisticated AI systems, as well as competitive pressures
for them to do so before anyone else. In the absence of any global
organization with the power to enforce a permanent moratorium on AI
development, disinclination and efforts at active prevention are likely
to be local and temporary.

It is worth emphasizing that, as with the arguments discussed in §2
above, the ubiquitous language of intelligence in discussions of the
Singularity Hypothesis is eliminable. As Chalmers shows,
singularity-type arguments can be constructed for any capability that is
either (i) self-amplifying in the sense that it obeys a version of the
proportionality thesis, or (ii) correlated with a self-amplifying
capability. For example, if the capability to design and train
deep-learning systems is self-amplifying and the capability to do
scientific research is correlated with the capability to design and
train deep-learning systems, there are singularity-style arguments that
we should expect to see an ``explosion'' of both capabilities after AI
reaches human-level competence at designing and training deep-learning
systems.

Given their radical conclusions, arguments for the Singularity
Hypothesis have always attracted controversy. For example, Modis (2012),
Plebe and Perconti (2012), and Thorstad (2022) argue on empirical
grounds that we should expect the rapid growth in the capabilities of
artificial systems observed in the past several decades to level off in
the near future rather than continue.\footnote{See also Vold and Harris
  (2021) for a brief overview.} Thorstad also engages directly with the
work of Chalmers and Bostrom, arguing that Chalmers does not
sufficiently motivate his proportionality thesis and that the reasons
Bostrom provides for thinking that recalcitrance will remain low in the
future are either unconvincing or do not motivate the idea of an
intelligence explosion.

In our view, neither the proponents of singularity-type arguments nor
their critics have made a fully convincing case. One important and
undertheorized question is whether levels of intelligence, or of any
given putative self-amplifying capability, have the right mathematical
structure for proportionality theses to make sense. To say that the
value of a function \emph{C}(\emph{t}) (say, our capacity to design
intelligent systems at time \emph{t}) increases in proportion to the
value of another function \emph{I}(\emph{t}) (say, the amount of
intelligence at our disposal at time \emph{t}) is to say that there is a
positive constant \emph{a} such that \emph{C}(\emph{t}) = \emph{a} *
\emph{I}(\emph{t}). The proportionality thesis therefore implies that
``the amount of intelligence at our disposal'' has something like the
structure of the real numbers, so that multiplication by a constant is
possible and the quantity \emph{a} * \emph{I}(\emph{t}) makes sense. But
if levels of intelligence have no more mathematical structure than that
of a total order, for example, we can't speak of \emph{a} *
\emph{I}(\emph{t}): we may be able to say of two amounts of intelligence
that the first is larger than the second, but not that the first is
\emph{a} times larger. Under these circumstances, then, it is not
possible to formulate a proportionality thesis.

A related question is whether proportionality theses, if they are
coherent, are plausible. While it seems clear that increases in a
system's intelligence will not be accompanied by \emph{reductions} in
its capacity to design intelligent systems or the optimization power it
is able to contribute to various problems, the claim that every increase
in intelligence is accompanied by a proportionate increase in design
capacity or optimization power is much stronger. Chalmers himself
considers the possibility of ``diminishing returns'' in design
capability as intelligence increases, which challenges premise 3 of his
argument. Why think that the proportionality thesis is more likely to be
true than the diminishing returns thesis?

While we sympathize with Thorstad's call for more evidence in support of
the proportionality thesis, we also think this thesis is less outlandish
than it may at first seem. As Chalmers formulates it, the
proportionality thesis is that increases in intelligence \emph{lead to}
proportionate increases in design capability. But the way in which this
leading-to occurs might be quite indirect. For example, increasing the
intelligence of a given system by 10\% might lead to a proportionate
increase in design capability by giving human engineers ideas about how
to improve the intelligence of a distinct system --- one better at
designing AI systems --- by 10\%. Nor do the increases posited by the
proportionality thesis need to happen immediately or be deterministic.
It could take years of chancy trial and error for a 10\% increase in the
intelligence of AI systems to translate into a 10\% improvement in their
design capabilities. And the proportionality thesis does not need to
hold in full generality to support Chalmers's argument: all that is
required is that it hold for a number of capability improvements
sufficient to achieve AI++ (Chalmers 2010, 22).

\section{Conclusion}

Are increasingly capable AI systems developed in the future likely to
seek power? Will progress in AI lead to the kind of recursive
improvement described by the Singularity Hypothesis? While some
theorists have reached strong conclusions in this area, these questions
strike us as unsettled. Continued philosophical research must be done to
assess the plausibility of arguments for catastrophic risk from AI
systems.

\section*{References}

{
\small

Baker, B., Kanitscheider, I., Markov, T., Wu, Y., Powell, G., McGrew, B.
and Mordatch, I. (2020). Emergent Tool Use From Multi-Agent
Autocurricula. \emph{International Conference on Learning
Representations 2020.}

Bales, A. (forthcoming). Will AI avoid exploitation? Artificial general
intelligence and expected utility theory. \emph{Philosophical Studies.}

Bostrom, N. (1998). How long before superintelligence?
\emph{International Journal of Future Studies}, 2. Reprinted in
\emph{Linguistic and Philosophical Investigations} 5: 11-30 (2006).

Bostrom, N. (2003). Ethical issues in advanced artificial
intelligence, in Smit, I. and Lasker, G. E. (eds.) \emph{Cognitive,
Emotive and Ethical Aspects of Decision Making in Humans and in
Artificial Intelligence}, Vol. 2. International Institute of Advanced
Studies in Systems Research and Cybernetics. Reprinted in Wallach, W.
and Asaro, P. (eds.) \emph{Machine Ethics and Robot Ethics}, Routledge,
2017, pp. 69--75.

Bostrom, N. (2012). The superintelligent will: Motivation and
instrumental rationality in advanced artificial agents. \emph{Minds and
Machines} 22, 71--85.

Bostrom, N. (2014). \emph{Superintelligence: Paths, Dangers,
Strategies}. Oxford University Press.

Bran, A.M., Cox, S., Schilter, O., Baldassari, C., White, A.D., and
Schwaller, P. (2023). \emph{ChemCrow: Augmenting large-language models
with chemistry tools.}
\href{https://arxiv.org/abs/2304.05376}{https://arxiv.org/abs/2304.05376}

Butlin, P., Long, R., et al. \emph{Consciousness in Artificial
Intelligence: Insights from the Science of Consciousness.}
https://arxiv.org/pdf/2308.08708.pdf

Carlsmith, J. (2022). \emph{Is Power-Seeking AI an Existential Risk?}
https://arxiv.org/abs/2206.13353

Carlsmith, J. (forthcoming). Existential risk from power-seeking AI. In
J. Barrett, H. Greaves \& D. Thorstad (eds.), \emph{Essays on
Longtermism}, OUP.

Chalmers, D. J. (2010). The Singularity: A Philosophical Analysis.
\emph{Journal of Consciousness Studies}, 17(9--10), 9--10.

Clark, J. and Amodei, D. (2016). Faulty Reward Functions in the Wild.
https://openai.com/research/faulty-reward-functions

Cotra, A. (2020). Forecasting TAI with Biological Anchors.
https://drive.google.com/drive/u/0/folders/ 15ArhEPZSTYU8f012bs6ehPS6-xmhtBPP.

D'Alessandro, W., Lloyd, H. and Sharadin, N. (2023). Large Language
Models and Biorisk. \emph{American Journal of Bioethics} 23 (10),
115-118.

Davidson, T. (2023). What a Compute-Centric Framework Says about Takeoff
Speeds. Open Philanthropy Report.
https://www.openphilanthropy.org/research/what-a-compute-centric-framework-says-about-takeoff-speeds/.

Drexler, K. E. (2019). \emph{Reframing Superintelligence: Comprehensive
AI Services as General Intelligence} (Technical Report \#2019-1). Future
of Humanity Institute.

Dreyfus, H. (1965). Alchemy and Artificial Intelligence. RAND
Corporation Technical Report.

Dreyfus, H. (1972). \emph{What Computers Can't Do}. Harper and Row.

Gallow, D. (forthcoming). Instrumental Divergence. \emph{Philosophical Studies}.
 
Gabriel, I. (2020). Artificial Intelligence, Values, and Alignment.
\emph{Mind and Machines.} 30(3), 411--437.

Goertzel, B. (2015). Superintelligence: Fears, Promises and Potentials.
Journal of Evolution \& Technology 25: 55--87.

Goldstein, S. and Kirk-Giannini, C. D. (2023). Language Agents Reduce
the Risk of Existential Catastrophe. \emph{AI \& Society.} Online First.

Good, I. J. (1951). Review of \emph{Calculating Instruments and
Machines} by D. R. Hartree. \emph{Journal of the Royal Statistical
Society Series A} 114: 106-107.

Good, I. J. (1959). Speculations on perceptrons and other automata. IBM
Research Rept. No. RC-116.

Good, I. J. (1962). The social implications of artificial intelligence.
In Good, I. J., Maybe, A. J. and Smith, J. M. (eds.) \emph{The Scientist
Speculates}, William Heinemann Ltd., pp. 192-198.

Good, I. J. (1965). Speculations concerning the first ultraintelligent
machine, in Alt, F. \& Rubinoff, M. (eds.) \emph{Advances in Computers},
vol 6, Academic Press.

Goodfellow, I., Bengio, Y., and Courville, A. (2016). \emph{Deep
Learning.} The MIT Press.

Grace, K., Stewart, H., Sandkühler, J.F., Thomas, S., Weinstein-Raun, B.
and Brauner, J. (2024). Thousands of AI authors on the future of AI.
\emph{AI Impacts},
https://aiimpacts.org/wp-content/uploads/2023/04/Thousands\_of\_AI\_authors\_on\_the\_future\_of\_AI.pdf.

Hanson, R. (2008). Economics of the singularity. \emph{IEEE Spectrum}
45(6): 37--43.

Hendrycks, D., Mazeika, M. and Woodside, T. (2023). \emph{An Overview of
Catastrophic AI Risks.} Center for AI Safety.
https://arxiv.org/pdf/2306.12001.pdf

Hubinger, E., van Merwijk, C., Mikulik, V., Skalse, J. and Garrabrant,
S. (2019). \emph{Risks from Learned Optimization in Advanced Machine
Learning Systems.} https://arxiv.org/abs/1906.01820

Jumper, J., Evans, R. et al. (2021). Highly accurate protein structure
prediction with AlphaFold. \emph{Nature.} 596, 583--589.

Karnofsky, H. (2022). AI could defeat all of us combined\emph{. Cold
Takes}. https://www.cold-takes.com/ai-could-defeat-all-of-us-combined

Klee, M. and Ramirez, N. M. 2023. AI Has Made the Israel-Hamas
Misinformation Epidemic Much, Much Worse. \emph{Rolling Stone}, October
27, 2023.

Krakovna, V., Uesato, J., Mikulik, V., Rahtz, M., Everitt, T., Kumar, R.
Kenton, Z., Leike, J. and Legg, S. (2020). Specification Gaming: The
Flip Side of AI Ingenuity.
https://www.deepmind.com/blog/specification-gaming-the-flip-side-of-ai-ingenuity.

Kurzweil, R. (2005). \emph{The Singularity is Near}. Viking.

Langosco, L., Koch, J., Sharkey, L.D., Pfau, J. and Krueger, D. (2022).
Goal misgeneralization in deep reinforcement learning. In
\emph{International Conference on Machine Learning}, Proceedings of
Machine Learning Research, pp. 12004--12019.

Loosemore, R. and Goertzel, B. (2012). Why an Intelligence Explosion is
Probable. In Eden, A. H., Moor, J. H., Søraker, J. H., and Steinhart, E.
(eds.) \emph{Singularity Hypotheses: A Scientific and Philosophical
Assessment}, Springer, pp. 83--96.

Modis, T. (2012). Why the Singularity Cannot Happen. In Eden, A. H.,
Moor, J. H., Søraker, J. H., and Steinhart, E. (eds.) \emph{Singularity
Hypotheses: A Scientific and Philosophical Assessment}, Springer, pp.
311--339.

Moravec, H. (1988). \emph{Mind Children: The Future of Robot and Human
Intelligence}. Harvard University Press.

Muehlhauser, L. and Salamon, A. (2012). Intelligence Explosion: Evidence
and Import. In Eden, A. H., Moor, J. H., Søraker, J. H., and Steinhart,
E. (eds.) \emph{Singularity Hypotheses: A Scientific and Philosophical
Assessment}, Springer, pp. 15--40.

Muehlhauser, L. and Helm, L. (2012). The Singularity and Machine Ethics.
In Eden, A. H., Moor, J. H., Søraker, J. H., and Steinhart, E. (eds.)
\emph{Singularity Hypotheses: A Scientific and Philosophical
Assessment}, Springer, pp. 101--126.

Müller, V. C., \& Cannon, M. (2021). Existential risk from AI and
orthogonality: Can we have it both ways? \emph{Ratio}, 35(1), 25--36.

Ngo, R., Chan, L. and Mindermann, S. (2023). \emph{The Alignment Problem
from a Deep Learning Perspective (v5).} https://arxiv.org/abs/2209.00626

OpenAI. (2023a). GPT-4 Technical Report.
https://arxiv.org/abs/2303.08774

OpenAI. (2023b). GPT-4 System Card.
https://cdn.openai.com/papers/gpt-4-system-card.pdf

Omohundro, S. (2007). \emph{The Nature of Self-Improving Artificial
Intelligence.}
https://selfawaresystems.files.wordpress.com/2008/01/nature\_of\_self\_improving\_ai.pdf

Omohundro, S. (2008). The basic AI drives. In P. Wang, B. Goertzel and
S. Franklin (eds.), \emph{Proceedings of the First Conference on
Artificial General Intelligence}, IOS Press.

Ord, T. (2020). \emph{The Precipice: Existential Risk and the Future of
Humanity}. Bloomsbury Publishing.

Pinker, S. (2015). Thinking Does Not Imply Subjugating. In Brockman, J.
(ed) \emph{What To Think About Machines That Think,} Harper Perennial,
pp. 5--8

Plebe, A. and Perconti, P. (2012). The Slowdown Hypothesis. In Eden, A.
H., Moor, J. H., Søraker, J. H., and Steinhart, E. (eds.)
\emph{Singularity Hypotheses: A Scientific and Philosophical
Assessment}, Springer, pp. 349--362.

Russell, S. (2019). \emph{Human Compatible: AI and the Problem of
Control.} Allen Lane.

Russell, S. and Norvig, P. (2021). \emph{Artificial Intelligence: A
Modern Approach (4th edition).} Pearson.

Salib, P. (Forthcoming). \emph{AI Will Not Want to Self-Improve}.
Lawfare Digital Social Contract Whitepapers.

Sanger, David E. and Steven Lee Myers. 2023. China Sows Disinformation
About Hawaii Fires Using New Techniques. \emph{New York Times}, Sep. 11
2023.

Sattarov, F. (2019). \emph{Power and Technology: A Philosophical and
Ethical Analysis.} Rowman \& Littlefield Publishers.

Schrittwieser, J., Antonoglou, I., Hubert, T., Simonyan, K., Sifre, L.,
Schmitt, S., Guez, A., Lockhart, E., Hassabis, D., Graepel, T.,
Lillicrap, T., and Silver, D. (2019). \emph{Mastering Atari, Go, Chess
and Shogi by Planning with a Learned Model}.
https://arxiv.org/abs/1911.08265

Schwitzgebel, E. and Garza\emph{,} M. (2015). A Defense of the Rights of
Artificial Intelligence. \emph{Midwest Studies in Philosophy.} 39(1),
98--119.

Schwitzgebel, E. and Garza\emph{,} M. (2020). Designing AI with Rights,
Consciousness, Self-Respect, and Freedom. In S. Liao (ed). \emph{Ethics
of Artificial Intelligence.} Oxford University Press.

Shah, R., Varma, V., Kumar, R., Phuong, M., Krakovna, V., Uesato, J. and
Kenton, Z. (2022). \emph{Goal Misgeneralization: Why Correct
Specifications Aren\textquotesingle t Enough For Correct Goals}.
https://arxiv.org/abs/2210.01790

Shulman, C. and Bostrom, N. (2021). Sharing the World with Digital
Minds. In S. Clarke, H. Zohny, \& J. Savulescu (eds.). \emph{Rethinking
Moral Status}. Oxford University Press.

Thornley, E. (2023). There Are No Coherence Theorems. \emph{The
Effective Altruism Forum.}
\href{https://forum.effectivealtruism.org/posts/FoRyordtA7LDoEhd7/there-are-no-coherence-theorems}{https://forum.effectivealtruism.org/posts/FoRyordtA7LDoEhd7/there-are-no-coherence-theorems}.

Thorstad, D. (2022). Against the Singularity Hypothesis. Global
Priorities Institute Working Paper No. 19-2022.

Thorstad, D. (2023). Exaggerating the Risks (Part 8: Carlsmith Wrap-Up).
\emph{Reflective Altruism.}
https://ineffectivealtruismblog.com/2023/06/03/exaggerating-the-risks-part-8-carlsmith-wrap-up/

Turner, A.M., Smith, L., Shah, R., Critch, A., Tadepalli, P. (2021).
Optimal Policies Tend to Seek Power. \emph{Advances in Neural
Information Processing Systems}, 35.

Turner, A.M., Tadepalli, P. (2022). Parametrically Retargetable
Decision-Makers Tend To Seek Power. \emph{Advances in Neural Information
Processing Systems}, 36.

Vinge, V. (1993). The coming technological singularity: How to survive
in the post-human era. \emph{Proceedings of Vision-21: Interdisciplinary
Science and Engineering in the Era of Cyberspace} (NASA Conference
Publication 10129): 11-22.

Vold, K. and Harris, D.R. (2021). How Does Artificial Intelligence Pose
an Existential Risk? In C. Véliz (ed.). \emph{Oxford Handbook of Digital
Ethics.} Oxford University Press.

Wang, L., Ma, C., Feng, X., Zhang, Z., Yang, H., Zhang, J., ... and Wen,
J. R. (2023). A survey on large language model based autonomous agents.
\href{https://arxiv.org/abs/2308.11432}{https://arxiv.org/abs/2308.11432}.

Wynroe, K., Atkinson, D., and Sevilla, J. (2023). Literature review of
transformative artificial intelligence timelines. \emph{Epoch AI},
https://epochai.org/blog/literature-review-of-transformative-artificial-intelligence-timelines.

Yampolskiy, R. (2015). Taxonomy of Pathways to Dangerous AI\emph{. AAAI
Workshop: AI, Ethics, and Society.}

Yudkowsky, E. (2008). Artificial Intelligence as a Positive and Negative
Factor in Global Risk. In Bostrom, N. and Ćirković, M. M. (eds.)
\emph{Global Catastrophic Risks}, Oxford University Press, pp. 308--345.

Yudkowsky, E. (2019). Coherent Decisions Imply Consistent Utilities.
\emph{LessWrong.}
https://www.lesswrong.com/posts/RQpNHSiWaXTvDxt6R/coherent-decisions-imply-consistent-utilities

Yudkowsky, E. (2023). Pausing AI Development Isn't Enough. We Need to
Shut it All Down. \emph{Time}, March 29, 2023.
https://time.com/6266923/ai-eliezer-yudkowsky-open-letter-not-enough/

Zwetsloot, R. and Dafoe, A. (2019). \emph{Thinking About Risks From AI:
Accidents, Misuse and Structure.} Lawfare.

}


\end{document}